\documentclass{aa}  

\usepackage{graphicx,natbib,url,twoopt}
\usepackage{txfonts}           
\usepackage{hyperref}                
\hypersetup{
    colorlinks=true,   
    urlcolor=blue,     
    linkcolor=red,     
}

\usepackage{pdfcomment}              
\usepackage{acronym}                 

\usepackage{subfigure}   
\usepackage{multirow}
\usepackage{amssymb,bm}
\usepackage{pdflscape}
\usepackage{color}
\usepackage{booktabs}
\usepackage{array}

\begin{document}

   \title{Systematics and accuracy of VLBI astrometry: What can be learned from a comparison with \textit{Gaia} Data Release 2}
   \titlerunning{Systematics and accuracy of ICRF}

   \author{N. Liu
          \inst{1,2}
          \and
          S. B. Lambert\inst{2}
          \and
          Z. Zhu\inst{1}
          \and
          J.-C. Liu\inst{1}
          }

   \institute{
   School of Astronomy and Space Science, 
         Key Laboratory of Modern Astronomy and Astrophysics (Ministry of Education), \\
         Nanjing University, 
         163 Xianlin Avenue, 210023 Nanjing, P. R. China \\
         \email{liuniu@smail.nju.edu.cn;zhuzi@nju.edu.cn}
         \and
          SYRTE, Observatoire de Paris, Université PSL, CNRS, Sorbonne Université, LNE, Paris, France\\
            \email{sebastien.lambert@obspm.fr}
             }

   \date{Received; accepted}

 
  \abstract
   {
       The third generation of the ICRF -- ICRF3 -- was published in 2018.
       This new fundamental catalog provides radio source positions measured independently at three bands: S/X, K, and X/Ka, 
       representing three independent radio celestial frames which altogether constitute together a multi-frequency ICRF.
   }
   {
       We aim to investigate the overall properties of the ICRF3 with the help of the \textit{Gaia} Data release 2 (\textit{Gaia} DR2).
       This could serve as an external check of the quality of the ICRF3.
   }
   {
       The radio source positions of the ICRF3 catalog were compared with the \textit{Gaia} DR2 positions of their optical counterparts at $G<18.7$.
       Their properties were analyzed in terms of the dependency of the quoted error on the number of observations, on the declination, and the global difference, the latter revealed by means of expansions in the vector spherical harmonics.
   }
   {
       The ICRF3 S/X-band catalog shows a more smooth dependency on the number of observations than the ICRF1 and ICRF2, while the K and X/Ka-band yield a dependency discrepancy at the number of observations of $\sim$50.
       The rotation of all ICRF catalogs show consistent results, except for the $X$-component of the X/Ka-band which arises from the positional error in the non-defining sources.
       No significant glides were found between the ICRF3 S/X-band component and \textit{Gaia} DR2.
       However, the K- and X/Ka-band frames show a dipolar deformation in $Y$-component of $\mathrm{+50\,\mu as}$ and several quadrupolar terms of $\mathrm{50\,\mu as}$ in an absolute sense.
       A significant glide along $Z$-axis exceeding $\mathrm{200~\mu as}$ in the X/Ka-band was also reported.
       These systematics in the ICRF catalog are shown to be less dependent on the limiting magnitude of the ${\it Gaia}$ sample when the number of common sources is sufficient ($>\,100$).
   }
   {
       The ICRF3 S/X-band catalog shows improved accuracy and systematics at the level of noise floor.
       But the zonal errors in the X/Ka-band should be noted, especially in the context of comparisons of multi-frequency positions for individual sources.
   }

   \keywords{Techniques : interferometric --
                Astrometry --
                reference systems
               }

   \maketitle
\section{Introduction}     \label{sec:intro}
The celestial reference frame (CRF) is a cornerstone of several scientific domains connected to astrometry and geosciences.
In its current realization by very long baseline interferometry (VLBI), the third generation of the International Celestial Reference Frame (ICRF3) is represented by precise positions of thousands of extragalactic radio sources \citep{2018IAU.....ICRF3}.
In contrast to its predecessors, which are the ICRF1 \citep{1998AJ....116..516M} and ICRF2 \citep{2015AJ....150...58F} achieved only at the S/X-band, the ICRF3 contains radio source positions at three bands: S/X, K, and X/Ka.
An accurate CRF opens the doors to millimeter-level geodetic measurements (e.g., Earth rotation), insights into the Earth internal structure \citep[e.g.,][]{2002JGRB..107.2068M}, and testing of General Relativity and alternative theories with VLBI \citep[][and references therein]{2016PhRvD..94l5030L}.
The arrival of the {\it Gaia} Celestial Reference Frame \citep[{\it Gaia}-CRF2,][]{2016A&A...595A...1G,2018A&A...616A..14G} - the first time that one can compare two independent CRF realizations with similar high accuracies - also allows comparisons between the positions of the common objects at several frequencies, bringing insights into the physics of quasars \citep[e.g.,][]{2017A&A...598L...1K}, reference frames \citep[e.g.,][]{2014A&A...570A.108L}, and complex structures made of one or several black holes \citep[e.g.,][]{2015A&A...578A..86R}.

Geodetic VLBI measures positions of thousands of radio sources since 1979.
\citet{2015AJ....150...58F} determined an internal precision of the ICRF2 and claimed a noise floor (best positional accuracy) of $\mathrm{40\,\mu as}$ and remarked that the stability of the ICRF axes is close to $\mathrm{10\,\mu as}$.
\citet{2018IAU.....ICRF3} claimed the noise floor of the ICRF3 to be of $\mathrm{30\,\mu as}$, which was later validated in \citet{2018A&A...620A.160L}.
However, no other detailed information has been revealed yet.

The ICRF3 started to serve as the fundamental catalogs for astrometry, geodesy, and navigation since the beginning of 2019.
This situation will remain consistent over the next several years until the release of the next generation of the ICRF.
It is, therefore, necessary to study its overall properties carefully, especially the consistence between CRFs at different wavelengths.
Such work would also provide hints on the link between ICRF and \textit{Gaia}-CRF in the next data release of the \textit{Gaia} mission.

This work aims to study the overall properties of the ICRF3 with the help of the \textit{Gaia} Data Release 2 \citep[\textit{Gaia} DR2,][]{2018A&A...616A...1G}.
It could serve as an external check of the quality of the ICRF3 to validate the improvements of the S/X-band frame and reveal possible systematic errors if there are any.
The three components of the ICRF3 catalog represent realizations of the ICRF3 at three bands, that is, they make up the ICRF3 altogether.
However in this work, we prefer to treat them as the independent CRFs.
Another intention of this work is to check the consistencies amongst three CRFs.
\section{Catalogs}     \label{sec:data}
In this work, we used the ICRF catalogs available at the Paris Observatory IERS ICRS Center\footnote{\url{http://hpiers.obspm.fr/icrs-pc/newwww/index.php}}.
Historical ICRF catalogs ICRF1 and ICRF2 were included in the comparison in order to justify the improvements made at the S/X-band.
Table~\ref{tab:icrf-stat} summarizes the statistical information of ICRFs.
The noise floor given in the first column represents the best position precision of individual sources, 
which is claimed to be $\mathrm{30\,\mu as}$ for ICRF3 S/X-band frame but not clear for ICRF3 K- and X/Ka-band frames.
The axis stability shown in the second column characterizes how stable the orientation of each reference axis is.
The last two columns give the number of defining sources in the northern and southern hemispheres, which are nearly the same for the ICRF3 catalogs (all three parts).
It indicates a better balance on the defining source number in the south and north of ICRF3 than ICRF2.
\begin{table*}[!htbp]
    \centering
    \caption{Statistics of the ICRFs.
    }
    \label{tab:icrf-stat}
    \begin{tabular}{ccccccccc}
        \toprule
        \toprule
        &Noise floor  &Axis stability  &\multicolumn{3}{c}{All sources}  &\multicolumn{3}{c}{Defining sources} \\
        \cmidrule(lr){4-6}  \cmidrule(lr){7-9}
        &$\mathrm{\mu as}$  &$\mathrm{\mu as}$  & All  &South  &North  & All  &South  &North\\
        \midrule
        ICRF1         &250  &20  &608   &266    &342   &212   &58   &154   \\
        ICRF2         &40   &10  &3414  &1383   &2031  &295   &133  &162   \\
        ICRF3 S/X     &30   &-   &4536  &1921   &2625  &303   &154  &149   \\
        ICRF3 K       &-    &-   &824   &373    &451   &193   &99   &94    \\
        ICRF3 X/Ka    &-    &-   &678   &361    &317   &176   &89   &87    \\
        \bottomrule
    \end{tabular}
    \tablefoot{
        A dash (`-') means that the corresponding value is not given or documented yet.
        The last 6 columns give the number of sources in each group for all sources and defining sources only.}
\end{table*}

We used the \texttt{gaiadr2.aux\_iers\_gdr2\_cross\_id} sample from \textit{Gaia} DR2 archive\footnote{\url{http://gea.esac.esa.int/archive/}} as the reference catalog.
This data set, might be called ICRF3-prototype subset, provides optical positions for 2820 ICRF sources.
It was used for aligning the \textit{Gaia}-CRF2 onto a prototype solution of the ICRF3 \citep{2018A&A...616A..14G}.
Even though more optical counterparts of ICRF sources could be found by cross-matching radio catalogs with the full quasar sample or the main catalog in the \textit{Gaia} DR2, we preferred to use this carefully identified catalog to avoid additional errors introduced by the cross-match process.
Since we were dealing with a global feature, we believed that adding a few more common objects in the comparison would not alter our results.

A good reference sample will benefit the catalog comparison and help to clearly see the systematics in the ICRF catalogs.
The accuracy of {\it Gaia} DR2 depends strongly on the magnitude, that is, brighter quasars tend to have positions with better formal uncertainties \citep{2018A&A...616A..14G}.
However, if we limit the sample to the bright quasars only, the sample will surely become smaller.
The trade-off between size and data quality is chosen to be at $G < 18.7$, where the sample size is 1288, about half of the ICRF3-prototype subset but the median positional uncertainty is twice better than the full population.
In the rest of this paper, for brevity, we will use the term ``\textit{Gaia} DR2'' to represent this subset, although it is a small fraction of the whole {\it Gaia} DR2 catalog.
The influence of the reference sample selection on the results is discussed in Sect.~\ref{sec:gaia-sample}.

In the compiling process of the ICRF3 catalog, the Galactic aberration effect has been modeled as a dipolar proper motion field toward the Galactic center with an amplitude of $\mathrm{5.8~\mu as~yr^{-1}}$.
As a result, the ICRF3 position is referred to the epoch J2015.0.
This is a new feature of the ICRF3.
The reference epoch for the \textit{Gaia} DR2 position is J2015.5, 0.5~yr later than the ICRF3.
The positional offset introduced by the Galactic aberration effect accumulated in half a year is about $\mathrm{3~\mu as}$, insignificant relative to the position formal errors of the \textit{Gaia} DR2 and ICRF3.
This small positional difference will be, therefore, ignored in the following analysis.

When considering the Galactic aberration effect, the source position in the ICRF1 and ICRF2 catalogs should be actually referred to their mean observing epoch.
The direct comparison of the ICRF1/ICRF2 position with the \textit{Gaia} DR2 would be influenced by the position offset resulted from the Galactic aberration effect accumulated in the time span between the mean observing epoch of individual source and J2015.5.
This issue could be solved by propagating the ICRF1/ICRF2 positions from their mean observing epoch to J2015.5 using the Galactic aberration induced proper motion, or by propagating the \textit{Gaia} DR2 position to the mean observing epoch for each source in the ICRF1/ICRF2 via their \textit{Gaia} DR2 proper motion.
However, we decided to keep the radio source position in the ICRF1/ICRF2 catalogs unchanged and directly compared it with the \textit{Gaia} DR2 one.
It could preserve the origin features of the ICRF1 and ICRF2 and avoid possible errors propagated from the proper motion system of the \textit{Gaia} DR2.
\section{Results and discussion}     \label{sec:result}
\subsection{Positional error}     \label{subsec:pos-err}
\begin{figure}[hbtp]
    \centering
    \includegraphics[width=\hsize]{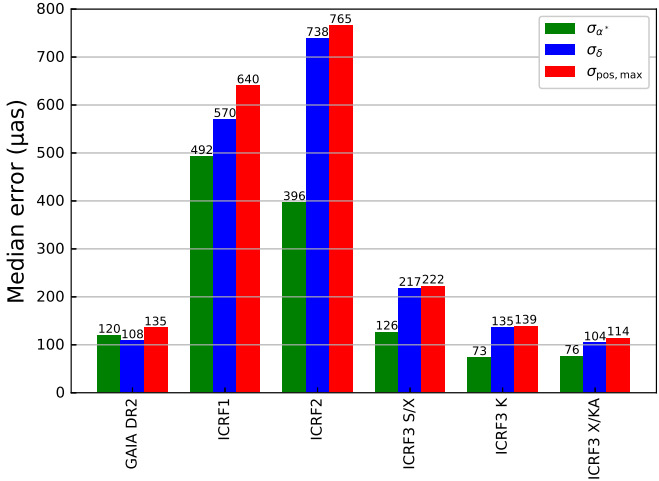}
    \caption[]{\label{fig:median-err}
        Formal errors of 1288 ICRF sources at $G<18.7$ in \textit{Gaia} DR2 and all sources in ICRF catalogs.
        The formal error in the right ascension is calculated as $\sigma_{\alpha^*} = \sigma_{\alpha}\cos\delta$.
    }
\end{figure}

The positional errors for ICRF catalogs and ${\rm Gaia}$ DR2 subset were characterized by median formal uncertainties in the right ascension (times cosine declination), declination, and along the major axis of error ellipse.
The semi-major axis of error ellipse, denoted as $\sigma_{\rm pos,max}$ as done in \citet{2018A&A...616A..14G} to characterize the overall positional error, was calculated with inclusion the correlation between the right ascension and declination.
Figure~\ref{fig:median-err} depicts these three quantities for the {\it Gaia} DR2 sample and ICRF catalogs.
For the {\it Gaia} DR2 sample, the median formal uncertainty is at the same level for right ascension and declination, which is of about $\mathrm{100\,\mu as}$. 
In contrast, all the ICRF catalogs present a better precision in the right ascension than in the declination; it is nearly twice better for the ICRF2, ICRF3 S/X-band, and K-band.
This is a typical feature of the VLBI measurements.
The  median formal error of ICRF2 declination is even larger than the ICRF1, possibly due to the inclusion of the so-called Very Long Baseline Array (VLBA) Calibrator sources (VCS sources) that were observed usually less than 100 times.
If we removed all VCS sources from the ICRF2 catalog, the median error for the remained sources would be reduced to $\mathrm{130\,\mu as}$ for right ascension and $\mathrm{194\,\mu as}$ for declination.
Thanks to the re-observations of these VCS sources \citep[VCS-II;][]{2016AJ....151..154G} which greatly improved the position precisions, the formal uncertainty of the ICRF3 S/X is nearly three times better than the ICRF2.
The ICRF3 S/X position shows a comparable (formal) uncertainty to the ${\rm Gaia}$ DR2 in the right ascension but nearly twice worse in the declination.
Another feature is that the VLBI position precision improves as the observing frequency goes higher.
The X/Ka-band catalog could be the most precise catalog in terms of the median formal error.

\begin{figure*}[hbtp]
    \centering
    \includegraphics[width=\hsize]{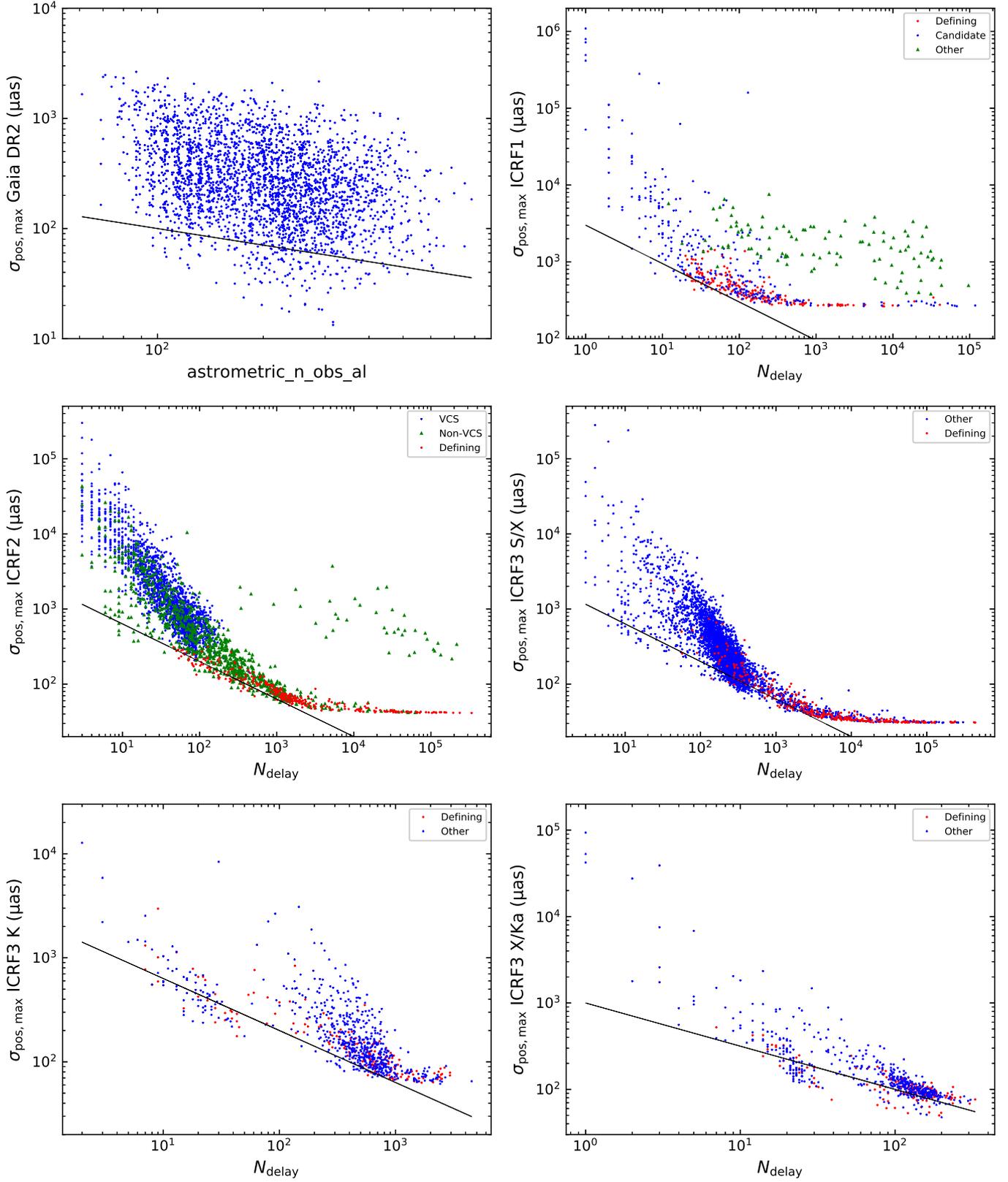}
    \caption[]{\label{fig:error-nobs}
        Positional errors as a function of the number of observations for 1288 ICRF sources at $G<18.7$ in \textit{Gaia} DR2 and all sources in ICRF catalogs.
        The positional error is computed as the semi-major axis of error ellipse using Eq.~(1) in \citet{2018A&A...616A..14G}.
        Sources in the ICRF catalogs are labeled by their category specified in each catalog.
        The black solid line indicates the power-slope from a Gaussian distribution.
        The number of observation is represented by the column \texttt{astrometric\_n\_obs\_al} in the \textit{Gaia} DR2 and the number of delays in the ICRF catalogs.
        
    }
\end{figure*}

The diagram in Fig.~\ref{fig:error-nobs} presents the distribution of the positional uncertainty as a function of the number of observations for the ICRFs and \textit{Gaia} DR2 sample. 
Sources in the ICRF catalogs are labeled by their category specified in each catalog.
The number of observations is represented by counting the number of used group delays, denoted as $N_{\it delay}$, in the ICRF catalog compilation and the along-scan individual CCD measurements (column `\texttt{astrometric\_n\_obs\_al}' in the {\it Gaia} archive) in the {\it Gaia} DR2.
We can find that the ${\it Gaia}$ observes quasars uniformly, at a frequency of about 100 to 400 times, which is attributed to its well-designed scanning law.
As a result, the positional errors of ICRF sources in the ${\it Gaia}$ DR2 are at a similar level of several tenths of mas.
On the contrary, the VLBI observation (delay) count ranges from single digit to over half a million at the S/X-band, leading to a less uniform accuracy.
The maximum observation per source is $\sim$4000 at K-band and $\sim$300 at X/Ka-band, at least two orders smaller than that at the S/X-band ($\sim$400\,000).
However, the best achieved (least) formal uncertainty at K-band and X/Ka-band is better than S/X-band.
This result justifies again the strength of observing at high frequencies.
The typical number of observations per source is about 100-400 for S/X-band, 400-600 for K-band, and 100-150 for X/Ka-band.

For ideal measurements, the formal error will fall down following a trend of 1/$\sqrt{N_{\rm obs}}$, where $N_{\rm obs}$ is the number of observations, as $N_{\rm obs}$ increases.
However, it is not the case for datasets shown in  Fig.~\ref{fig:error-nobs}.
All the ICRF catalogs show a non-Gaussian distribution at the rich-observation end.
This non-Gaussianity is caused by (i) correlated errors that become dominant as the thermal error ($1/N_{\rm obs}^2$) becomes lower for large $N_{\rm obs}$, and (ii) the inflation of the formal error.
The horizontal tail of the ICRF1, ICRF2, and ICRF3 S/X catalogs results from adjustment of a noise floor that is $\mathrm{250\,\mu as}$, $\mathrm{40\,\mu as}$, and $\mathrm{30\,\mu as}$, respectively \citep{1998AJ....116..516M,2015AJ....150...58F,2018IAU.....ICRF3}.

There are two features that need to be pointed out.
One is that some well-observed sources deviate from the main trend and yield bad positional precisions in the ICRF1 (``Other'' source) and ICRF2 (``Non-VCS'' sources, mostly known as 39 special-handling sources).
But such an inconsistency disappears in the ICRF3 S/X-band catalog which yields a smooth relation.
It could be regarded as an improvement of the ICRF3 S/X-band besides the reduce of the noise floor.
Another feature is that a clear dependency break is found in the ICRF3 K- and X/Ka-band, which happens at the number of observations of $\sim$50 for K-band and $\sim$40 for X/Ka-band.
    This discrepancy leads to a formal uncertainty ``jump'' and the dependency of the formal uncertainty could be better represented by two power-law slopes.
    Examination of the data reveals that sources at $N_{\rm delay}~<~50$ mostly fall in the region of $\delta\,<\,-40^\circ$, where the dominated network at K-band, the VLBA network, cannot cover.
    Therefore, it demonstrates a clear effect of K-band observing network geometry.
    Similar results can be found for X/Ka-band, where the bulk of sources with a number of observations less than 40 locates in the far-southern hemisphere and is hard to be observed by the two main baselines, from Goldstone (California) to Robledo (Spain) and from Goldstone to Tidbinbilla (Asutralia), in the X/Ka frequency.
    As a result, observations in the south at K- and X/Ka-band should be strengthened for the sake of reducing the north-south asymmetry in observations.
    With more uniform observations, the K-band and X/Ka-band will surely be improved, regarding the case of S/X-band.
\subsection{Declination-dependent error}     \label{subsec:dec-dep error}
\begin{figure}[htbp]
    \centering
    \includegraphics[width=\hsize]{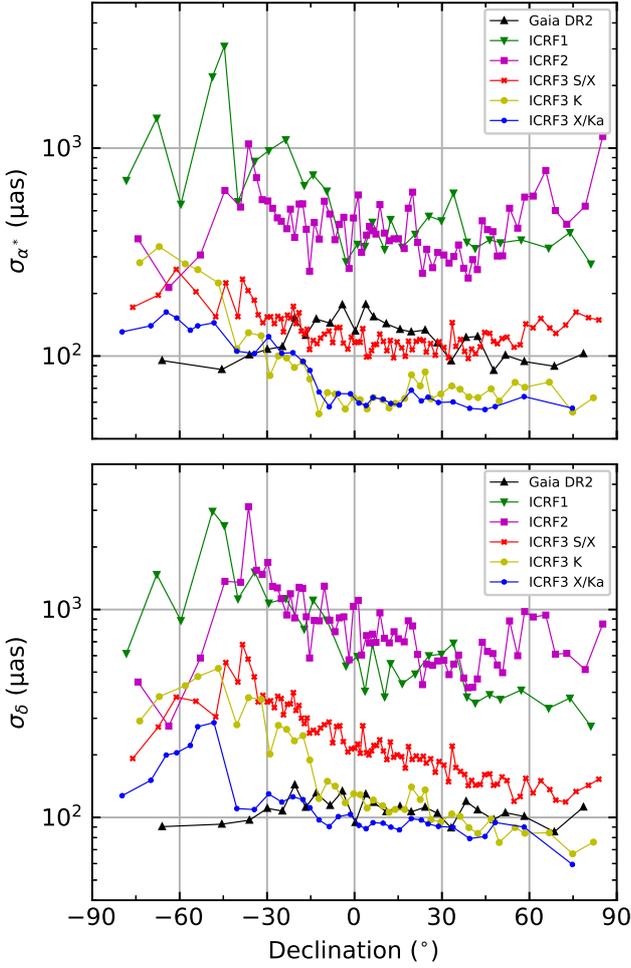}
    \caption[]{\label{fig:pos-err-vs-dec}
        Formal errors in the right ascension (times cosine declination, $top$) and declination ($bottom$) as a function of declination for the ICRF sources at $G<18.7$ in \textit{Gaia} DR2 and all sources in ICRF catalogs.
        In the right ascension the formal error is calculated as $\sigma_{\alpha^*} = \sigma_{\alpha}\cos\delta$.
        The shown formal error is the median of bins by every certain number of sources for the sake of clarity.
        The bin size was chosen to be 20 for the ICRF1, ICRF3 K-, and X/Ka-band, along with 50 for other catalogs.
    }
\end{figure}

As mentioned in Sect.~\ref{subsec:pos-err}, the south-north asymmetry in the formal uncertainty is a typical VLBI feature.
Thus we investigated the dependency of the position formal uncertainty in the ICRF and {\it Gaia} data on the declination.
The sources were sorted according to the declination from south to north and binned by every certain number of sources.
Figure~\ref{fig:pos-err-vs-dec} demonstrates the median formal errors in each bin as a function of declination.
The bin size was set as 20 for the ICRF1, ICRF3 K-band, and X/Ka-band, along with 50 for other catalogs.
The {\it Gaia} DR2 formal errors are generally homogeneous over the declination, of about 100~${\rm \mu}$as in both the right ascension and declination, except small increases near the equator.
On the other hand, all the ICRF catalogs show a similar declination-dependency: a bump (worst precision) appears at about $-$45$^\circ$ and the formal error decreases as going north until 30$^\circ$ then increases slightly.
This feature is more clear as seen from the formal error in the declination.
The decreasing dependency shown in the ICRF1/ICRF2 becomes smoother in the three catalogs of ICRF3, especially in the right ascension.
For sources at $\delta~>~-$15$^\circ$ in the ICRF3 catalogs at all frequencies, there is generally no declination dependency of the formal uncertainty in the right ascension component on the declination.
However, declination formal error of ICRF3 catalogs still show a decreasing trend on the declination, with an exception of ICRF3 X/Ka showing a smooth dependency in the north of $-$45$^\circ$.
This deficiency of ICRF3 three components, again, reflects the weakness of the VLBI network geometry in the south.

In principle, the {\it Gaia} scanning-law is not expected to generate a declination-dependent error or declination systematic error.
Figure~\ref{fig:pos-err-vs-dec} already demonstrates that {\it Gaia} error is homogeneous over the declination.
If only looking at the sources in each ICRF catalog in common with {\it Gaia} sample, we find similar relations of positional uncertainty on the declination for ICRF and {\it Gaia} samples shown in Fig.~\ref{fig:pos-err-vs-dec}.
As a result, we can compare the ICRF positions with respect to the {\it Gaia} DR2 positions in order to reveal possible declination-dependent (zonal) errors in the ICRF catalogs.

\begin{figure}[hbtp]
    \centering
    \includegraphics[width=\hsize]{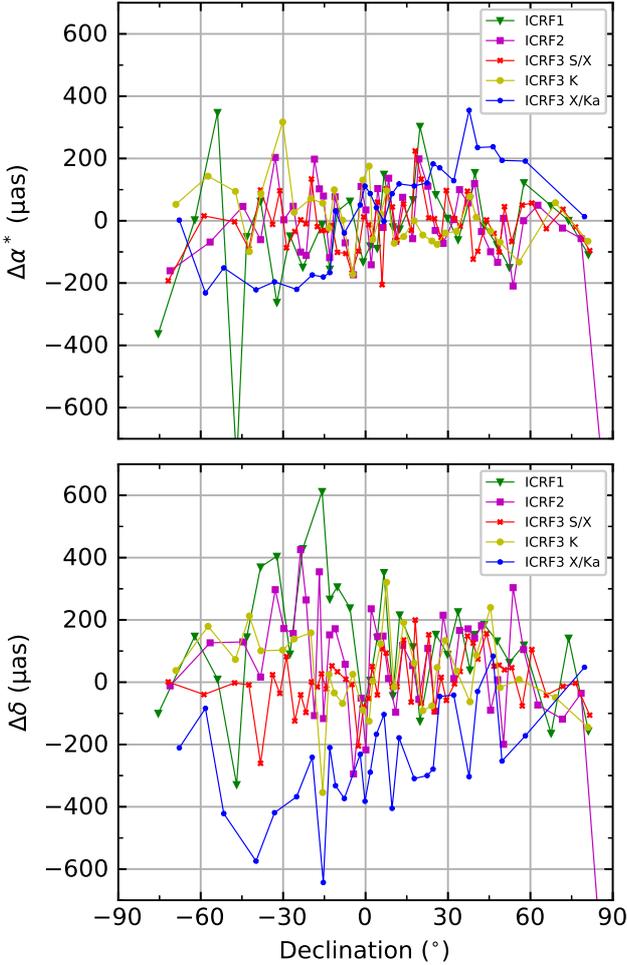}
    \caption[]{\label{fig:pos-diff-vs-dec}
        Differences in the right ascension (times cosine declination, $top$) and declination ($bottom$) as a function of declination for common sources between the \textit{Gaia} DR2 subset and ICRF catalogs.
        The right ascension difference is calculated as $\Delta\alpha^* = \Delta\alpha\cos\delta$.
        The shown difference is the median of bins by every certain number of sources for the sake of clarity.
        The bin size was chosen to be 10 for the ICRF1, ICRF3 K-, and X/Ka-band, along with 25 for other catalogs.
    }
\end{figure}

Figure~\ref{fig:pos-diff-vs-dec} illustrates the median differences in the right ascension and declination of ICRF catalogs with respect to the {\it Gaia} DR2 sample for common sources.
Sources were sorted and binned by the same way as in Fig.~\ref{fig:pos-err-vs-dec}, but different bin size was chosen; it was set to be 10 for the ICRF1, ICRF3 K-band, and X/Ka-band, along with 25 for other catalogs.
Some data points of ICRF1 and ICRF2 are beyond the frame but all less than 1\,mas.
On the right ascension, all the ICRF catalogs show consistent results within 200~$\mathrm{\mu as}$ with the {\it Gaia} DR2 except the ICRF3 X/Ka-band; it yields a sinusoidal-like pattern with an amplitude of approximately $200~\mathrm{\mu as}$.
On the contrary, the ICRF declinations vary a lot.
A positive bias can be found in the ICRF1 and ICRF2 catalogs in the southern hemisphere, as reported in previous studies \citep[e.g.,][]{2016A&A...595A...5M,2018A&A...609A..19L,2018AJ....155..229F}.
The agreement between the ICRF3 S/X and {\it Gaia} DR2 declinations is about 200~$\mathrm{\mu as}$ but no pattern could be found.
The ICRF3 K-band catalog also presents a nearly random difference with respect to {\it Gaia} solution.
These high consistencies amongst {\it Gaia} DR2, ICRF3 S/X-band, and K-band are satisfactory and justify the quality of all these three datasets.
However, the most obvious deviation from the {\it Gaia} DR2 comes from the ICRF3 X/Ka-band catalog; it shows a negative offset as large as about $\mathrm{\sim600~\mu as}$ at the regions around $\delta=-45^\circ$ and $\delta=-15^\circ$.
This anomalous behavior is possibly due to the weak southern geometry of the X/Ka-band network, indicating conspicuous zonal errors in the ICRF3 X/Ka-band frame.
Therefore, it is necessary to use a more detailed numerical analysis to quantify this systematics.

\subsection{Systematics}     \label{sec:systematics}
The systematical difference shown in Fig.~\ref{fig:pos-diff-vs-dec} can be modeled by a set of vector spherical harmonics \citep[VSH;][]{2012A&A...547A..59M}.
Here we use the VSH of the first two orders to represent the global systematical differences, as described by the following equations:
\begin{equation}
\label{eq:VSH}
\begin{array}{lll}
\Delta\alpha^*  &= &-R_1\cos\alpha\sin\delta  - R_2\sin\alpha\sin\delta + R_3\cos\delta \\
                &  &-D_1\sin\alpha            + D_2\cos\alpha \\
                &  &+M_{2,0}\sin 2\delta  \\
                &  &-\cos 2\delta~(M_{2,1}^{\rm Re}\cos  \alpha -  M_{2,1}^{\rm Im}\sin  \alpha) \\
                &  &+\sin  \delta~(E_{2,1}^{\rm Re}\sin  \alpha +  E_{2,1}^{\rm Im}\cos  \alpha) \\
                &  &-\sin 2\delta~(M_{2,2}^{\rm Re}\cos 2\alpha -  M_{2,2}^{\rm Im}\sin 2\alpha) \\
                &  &-2\cos \delta~(E_{2,2}^{\rm Re}\sin 2\alpha +  E_{2,2}^{\rm Im}\cos 2\alpha), \\
\Delta\delta    &= &+R_1\sin\alpha            - R_2\cos\alpha \\
                &  &-D_1\cos\alpha\sin\delta  - D_2\sin\alpha\sin\delta + D_3\cos\delta \\
                &  &+E_{2,0}\sin 2\delta  \\ 
                &  &-\sin  \delta~(M_{2,1}^{\rm Re}\sin  \alpha +  M_{2,1}^{\rm Im}\cos  \alpha) \\
                &  &-\cos 2\delta~(E_{2,1}^{\rm Re}\cos  \alpha -  E_{2,1}^{\rm Im}\sin  \alpha) \\
                &  &+2\cos \delta~(M_{2,2}^{\rm Re}\sin 2\alpha +  M_{2,2}^{\rm Im}\cos 2\alpha) \\
                &  &-\sin 2\delta~(E_{2,2}^{\rm Re}\cos 2\alpha -  E_{2,2}^{\rm Im}\sin 2\alpha), \\
\end{array}
\end{equation}
where $\Delta\alpha^* = \Delta\alpha\cos\delta$.
The first degree harmonics consists of a rotation vector $\bm{R} = (R_1, R_2, R_3)^{\rm T}$ and a glide vector $\bm{D} = (D_1, D_2, D_3)^{\rm T}$.
The rotation vector $\bm{R}$ characterizes the orientation between celestial frames, that is, how well these frames share the same reference axes.
The glide vector $\bm{D}$, on the other hand, reveals the dipolar deformation or zonal errors in the celestial frame.
As for the second degree, $E$ and $M$ stands for the VSH functions of electric and magnetic types, respectively.
These terms, also called quadrupole terms, are usually associated with the zonal errors in the catalog.

\begin{figure}[hbtp]
    \centering
    \includegraphics[width=\hsize]{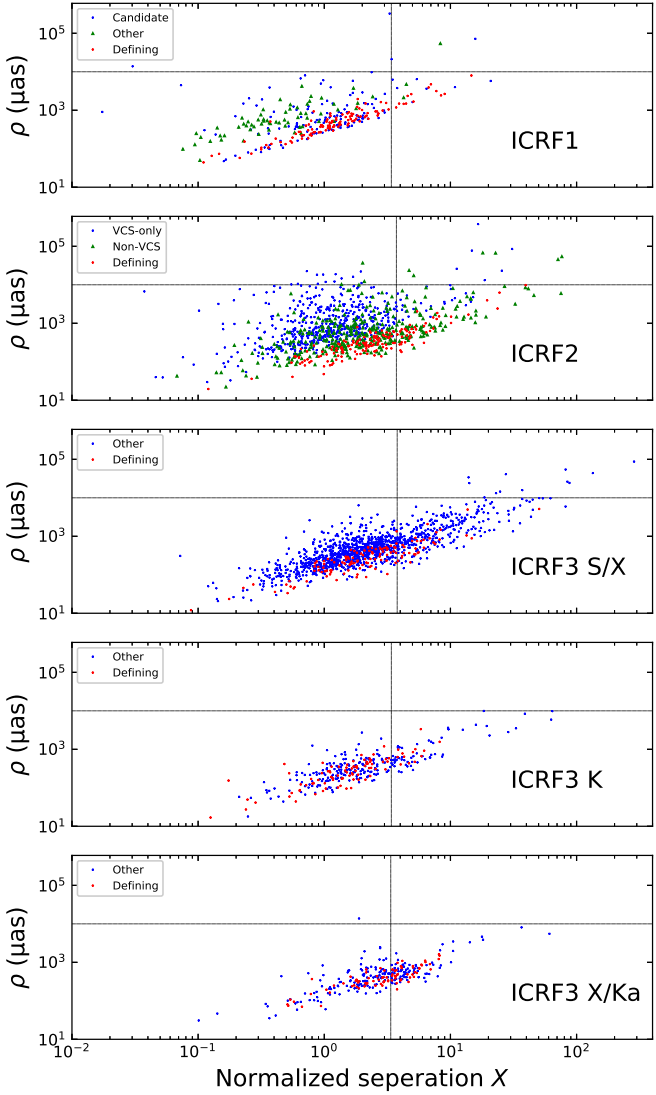}
    \caption[]{\label{fig:gaia-vlbi-separation}
        Angular separation and normalized separation between the \textit{Gaia} DR2 and ICRF catalogs for common sources at $G<18.7$.
        The horizontal gray line and vertical one indicates the upper limit of 10\,mas on the angular separation and of $X_0$ on the normalized separation, respectively.
    }
\end{figure}

Before addressing the systematics, we needed to remove outliers, say, the radio sources with a significant position offset between VLBI and {\it Gaia} that cannot be explained by its formal error.
These large discrepancies could result from the underestimated formal errors, misidentification, or genuine core-shift effect \citep[e.g.,][]{2019ApJ...873..132M,2017A&A...598L...1K}.
To remove these sources, we used the angular separation $\rho$ and $X$-statistics in \citet{2016A&A...595A...5M}.
$X$-statistic is a normalized separation considering not only the formal errors in the right ascension and declination but also the covariance between them in both catalogs.
The distribution of these two separation quantities is presented in Fig.~\ref{fig:gaia-vlbi-separation}.
For an ideally Gaussian noise, $X$ is supposed to follow a Rayleigh distribution.
The number of outliers for a sample of $N$ sources exceeds one when $X$ is greater than $X_0$, where $X_0 = \sqrt{2\ln N}$.
An additional cut-off threshold of 10\,mas was imposed on the angular separation.
The selection criteria can be summarized by the following equation as
\begin{equation}
\centering
\label{eq:constraint}
X \le X_0, \quad \rho \le 10~{\rm mas}.
\end{equation}
Finally, we obtained a ``clean'' sample, upon which we determined the coefficients of the VSH via a least square fit.
Table~\ref{tab:icrf-gdr2} provides the value of $X_0$ and number of common sources as well as the size of the clean sample between ICRF catalogs and {\it Gaia} sample.
The full covariance information for each source were used in the fit process, as done in \citet{2018A&A...609A..19L}.
These outliers are of interest of radio-optical offset studies and we address them in the Sect.\ref{sec:separation}.

\begin{figure}[hbtp]
    \centering
    \includegraphics[width=\hsize]{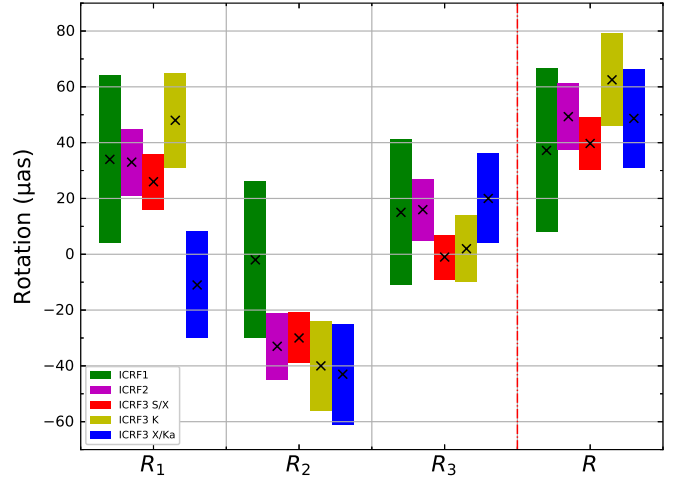}
    \caption[]{\label{fig:rotation}
        Rotation parameters of the ICRF positions with respect to the \textit{Gaia} DR2 position using common sources at $G<18.7$.
        $R$ is the module of rotation vector $\bm{R}$ calculated as $R=\sqrt{R_1^2+R_2^2+R_3^2}$.
        The black cross indicates the estimate from the fit while the bar marks the confidence interval of 1-$\sigma$.
    }
\end{figure}

\begin{table}[!htbp]
    \centering
    \caption{Statistics of the least-square fit.
    }
    \label{tab:icrf-gdr2}
    \begin{tabular}{cccccc}
        \toprule
        \toprule
        &$N$  &$X_0$  &$N^\prime$  &pre-fit $\chi^2$  &post-fit $\chi^2$ \\
        \midrule
        ICRF1         &326    &3.40   &298   &0.77   &0.67 \\
        ICRF2         &1052   &3.73   &876   &1.41   &1.35 \\
        ICRF3 S/X     &1288   &3.78   &871   &1.82   &1.80 \\
        ICRF3 K       &326    &3.40   &255   &1.47   &1.38 \\
        ICRF3 X/Ka    &286    &3.36   &168   &2.17   &1.31 \\
        \bottomrule& 
    \end{tabular}
    \tablefoot{$N$ stands for the number of common sources between ICRF catalogs and \textit{Gaia} sample while $N^\prime$ is the size of clean sample. 
        $X_0$ is computed as $X_0 = \sqrt{2\ln N}$.
        The last two columns give the pre- and  post-fit reduced-$\chi^2$.
    }
\end{table}
The rotation components between the $\rm Gaia$ DR2 and ICRF catalogs, demonstrated in Fig.~\ref{fig:rotation}, are generally at a level of 30-40~$\mathrm{\mu as}$.
This result is consistent with the claimed alignment accuracy of the $\it Gaia$-CRF2 \citep{2018A&A...616A..14G}.
Amongst ICRF catalogs, the rotation parameters are consistent within their formal errors, except for the $R_1$ of the ICRF3 X/Ka-band.
It means that the axial direction of the X/Ka-band frame deviates from that of S/X- and K-band frames at an angle of about $\mathrm{30~\mu as}$, which significantly exceeds the axial stability of ICRF ($\mathrm{10~\mu as}$ for the ICRF2).
However, if we compared directly the ICRF3 X/Ka-band catalog with the ICRF3 S/X-band catalog using exclusively the ICRF3 defining sources, the rotation angle around the $X$-axis would be smaller than $\mathrm{10\,\mu as}$.
This result gives us a hint that such an inconsistency would come from positional errors of the non-defining sources in the ICRF3 X/Ka-band catalog and should not be omitted in the comparison between the VLBI X/Ka-band and \textit{Gaia} positions.

Contrary to the rotation, the glide terms are significant relative to their formal errors, except for the X-component.
With respect to the \textit{Gaia} DR2, the ICRF1 shows the largest discrepancy, exceeding $\mathrm{100~\mu as}$, while the ICRF3 S/X frame shows the smallest.
The glide between the \textit{Gaia} DR2 and the ICRF3 S/X-band is consistent within the formal error.
This result shows that the \textit{Gaia}-CRF2 and the ICRF3 S/X band frame are highly consistent.
Assuming that the \textit{Gaia}-CRF2 is nearly free of declination-dependent systematics, it indicates that the dipolar deformation previously found in the ICRF2 has decreased significantly.
This achievement should be credited to the recent efforts to enhance the observations in the south \citep{2014ivs..confE...1J}, re-observations of the VCS-sources \citep{2016AJ....151..154G}, and the modeling of the Galactic aberration effect.
The significant glide in the ICRF1 and ICRF2 partly results from the zonal errors \citep[e.g.,][]{2018AJ....155..229F} and partly due to the unmodeled Galactic aberration effect \citep{2018IAU.....ICRF3}, both leading to the frame deformation.
A noticeable glide component over $\mathrm{\sim 200~\mu as}$ along the Z-axis appears in the ICRF3 X/Ka-band, revealing obvious zonal errors in the X/Ka-band frame as found in the Fig.~\ref{fig:pos-diff-vs-dec}.
The K- and X/Ka- band frames also present a similar dipolar deformation of $\mathrm{\sim 50~\mu as}$ along the Y-axis.

In terms of the quadrupolar parameters, the ICRF catalogs and the \textit{Gaia} DR2 agree on the level of $\mathrm{\sim 50~\mu as}$, especially for the ICRF3 S/X-band with all coefficients below $\mathrm{20~\mu as}$ except for the $E_{20}$.
Again, the ICRF3 X/Ka-band is an exception.
Two terms, $E_{20}$ and $M_{20}$, exceed $\mathrm{100~\mu as}$ and should be interpreted as zonal errors in the X/Ka-band frame.
Special attentions should also be given to terms that are above $\mathrm{50\,\mu as}$, such as $E^I_{21}$, $M_{21}^R$, and $M_{21}^I$ for the X/Ka-band, and $M_{21}^R$ for the K-band frame.
These terms should be examined carefully in the next release of the K- and X/Ka-band frames.

Based on the analysis of VSH technique, we found that the global agreement between ICRF3 S/X-band subset and {\it Gaia} sample is at the level of $\mathrm{30\,\mu as}$, and it is around $\mathrm{50\,\mu as}$ for ICRF3 K-band frame.
These satisfactory results indicate that these three realizations of the International Celestial Reference System (ICRS) agree well with the others. 
However, the significant dipolar and quadruploar terms found in the ICRF3 X/Ka-band datum with referred to the {\it Gaia} solution suggest severe zonal errors in the X/Ka-band frame.
\citet{2018IAU.....ICRF3} pointed out these zonal errors in X/Ka-band frame using S/X-band frame as the reference, which is similar to our results.
These zonal errors are likely due to 
(i) the small number of sources and also small number of observations for each sources;
(ii) the weak geometry (four antennas) of the X/Ka-band network.
Due to limited data on the baseline from Argentina to California, the X/Ka-band frame is vulnerable to the declination-dependent zonal errors.
At this stage, this emergent network has not reached its maturity.
Several plans to correct this deficiency were proposed in \citet{2018ivs..meetE...1D}.
Likely, the accuracy of the X/Ka-band frame will become better as comparable to the S/X- and K-band frames when the network has a better coverage and accumulates more observations.
\begin{figure}[hbtp]
    \centering
    \includegraphics[width=\hsize]{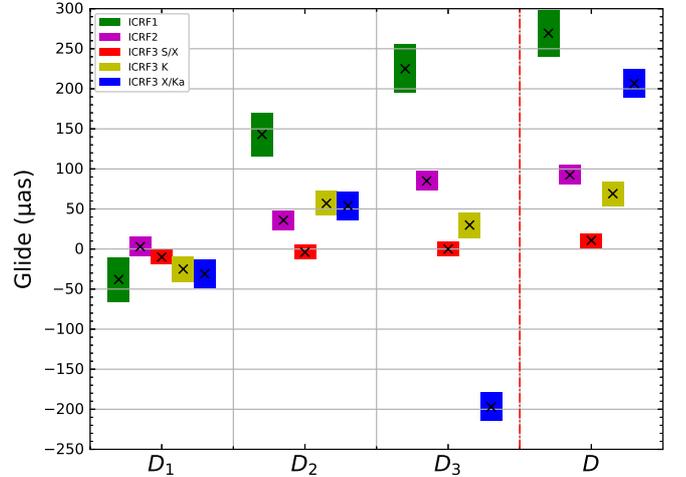}
    \caption[]{\label{fig:glide}
        Glide parameters of the ICRF positions with respect to the \textit{Gaia} DR2 position using common sources at $G<18.7$.
        $D$ is the module of glide vector $\bm{D}$ calculated as $D=\sqrt{D_1^2+D_2^2+D_3^2}$.
        The black cross indicates the estimate from the fit while the bar marks the confidence interval of 1-$\sigma$.
    }
\end{figure}

\begin{figure}[hbtp]
    \centering
    \includegraphics[width=\hsize]{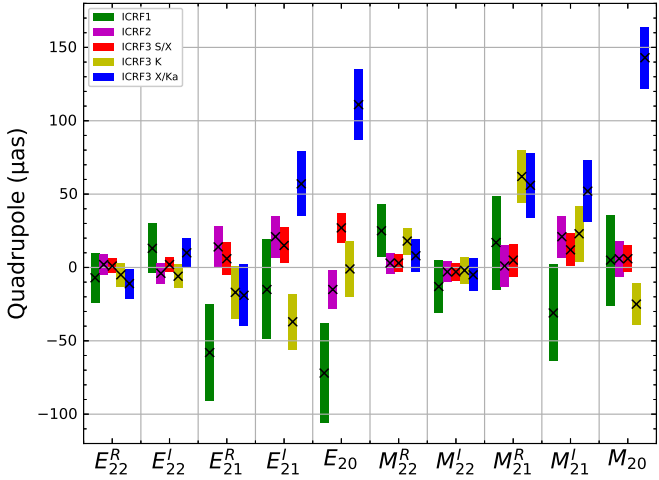}
    \caption[]{\label{fig:quadrupole}
        Quadrupole parameters of the ICRF positions with respect to the \textit{Gaia} DR2 position using common sources at $G<18.7$.
        The black cross indicates the estimate from the fit while the bar marks the confidence interval of 1-$\sigma$.}
\end{figure}
\subsection{Influence of limiting magnitude of {\it Gaia} sample}     \label{sec:gaia-sample}
\begin{figure}[hbtp]
    \centering
    \includegraphics[width=\hsize]{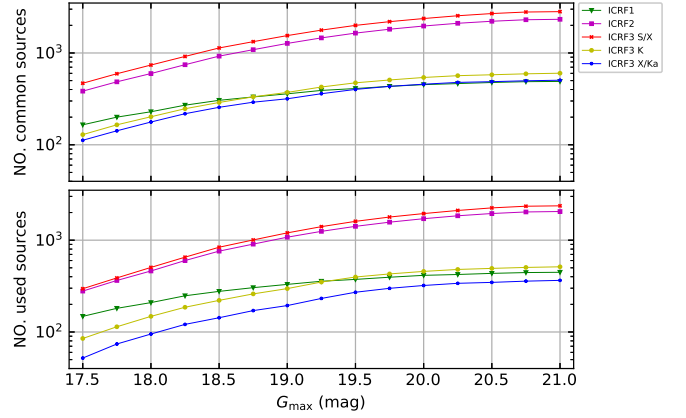}
    \caption[]{\label{fig:nb-sou-on-g}
        Distribution of numbers of all common sources between ICRF catalogs and the \textit{Gaia} DR2 ICRF3-prototype subset ($top$) and used sources in the fit ($bottom$) as a function of the limiting $G$-magnitude of {\it Gaia} sample.
    }
\end{figure}

As mentioned in Sect.~\ref{sec:data}, the formal uncertainty of {\it Gaia} data depends strongly on the magnitude.
As a result, we used only the bright sources ($G<18.7$) in the {\it Gaia} DR2 ICRF3-prototype subset.
To investigate the effect of limiting magnitude on the comparison,  we used {\it Gaia} subsets with different limiting $G$-magnitudes, labeled as $G_{\rm max}$, and determined the VSH parameters between ICRF and {\it Gaia} positions for common sources.
We chose a set of value in the range of 17.5-21.0 with a step of 0.5 for $G_{\rm max}$ and kept exclusively sources brighter than $G_{\rm max}$ in the sample.
The starting point was set as 17.5 in order to permit at least 100 common sources between ICRF catalogs and {\it Gaia} sample.
The VSH parameters were obtained following the same procedures in the Sect.~\ref{sec:systematics}.
Figure~\ref{fig:nb-sou-on-g} depicts the distribution of number of common sources and used sources in the fit as a function of the limiting $G$-magnitude.

\begin{figure}[hbtp]
    \centering
    \includegraphics[width=\hsize]{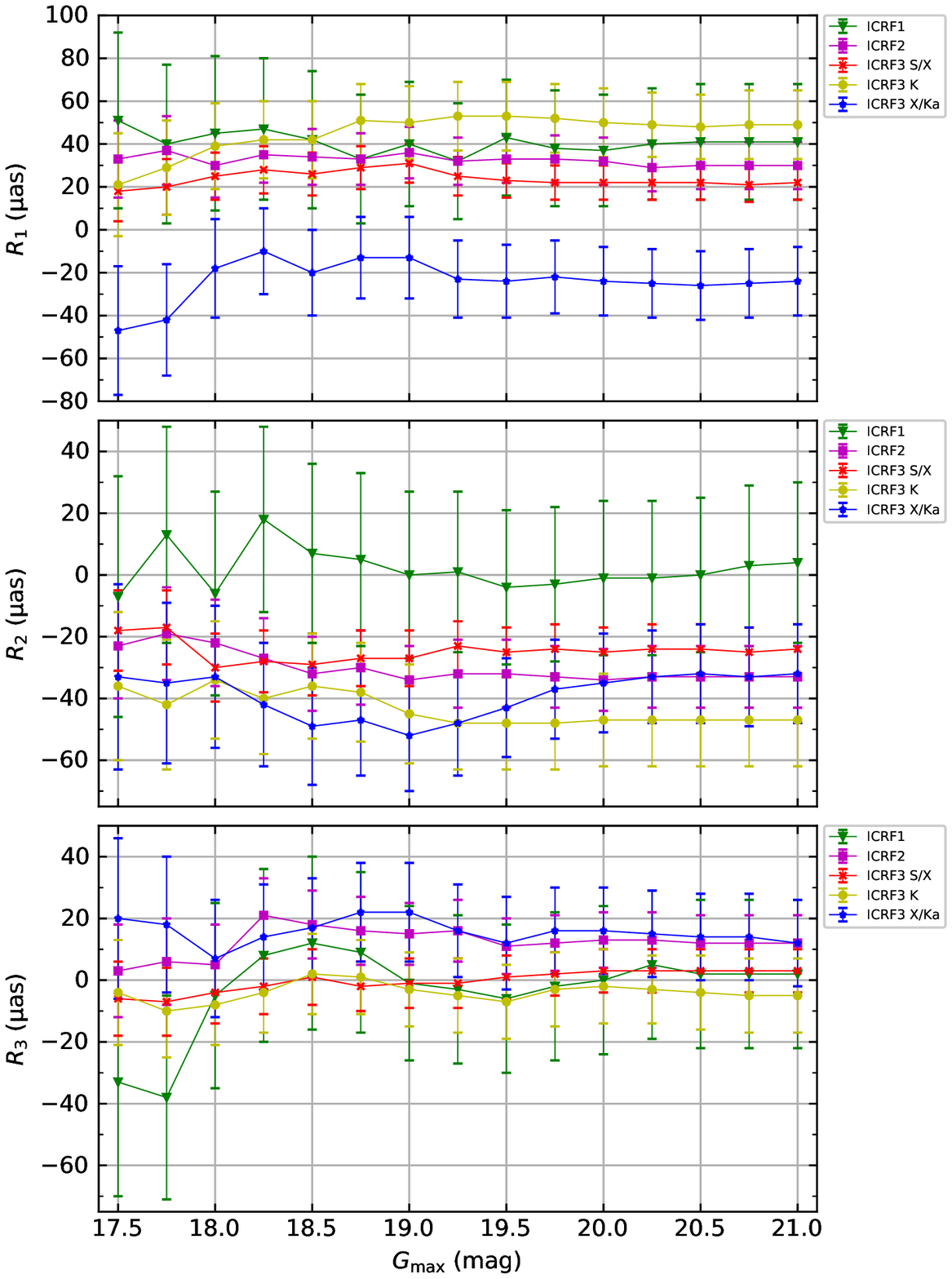}
    \caption[]{\label{fig:rotation-on-g}
        Rotation parameters of the ICRF positions with respect to the \textit{Gaia} DR2 position as a function of the limiting $G$-magnitude of {\it Gaia} sample.
        The error bar indicates the confidence interval of 1-$\sigma$.}
\end{figure}

\begin{figure}[hbtp]
    \centering
    \includegraphics[width=\hsize]{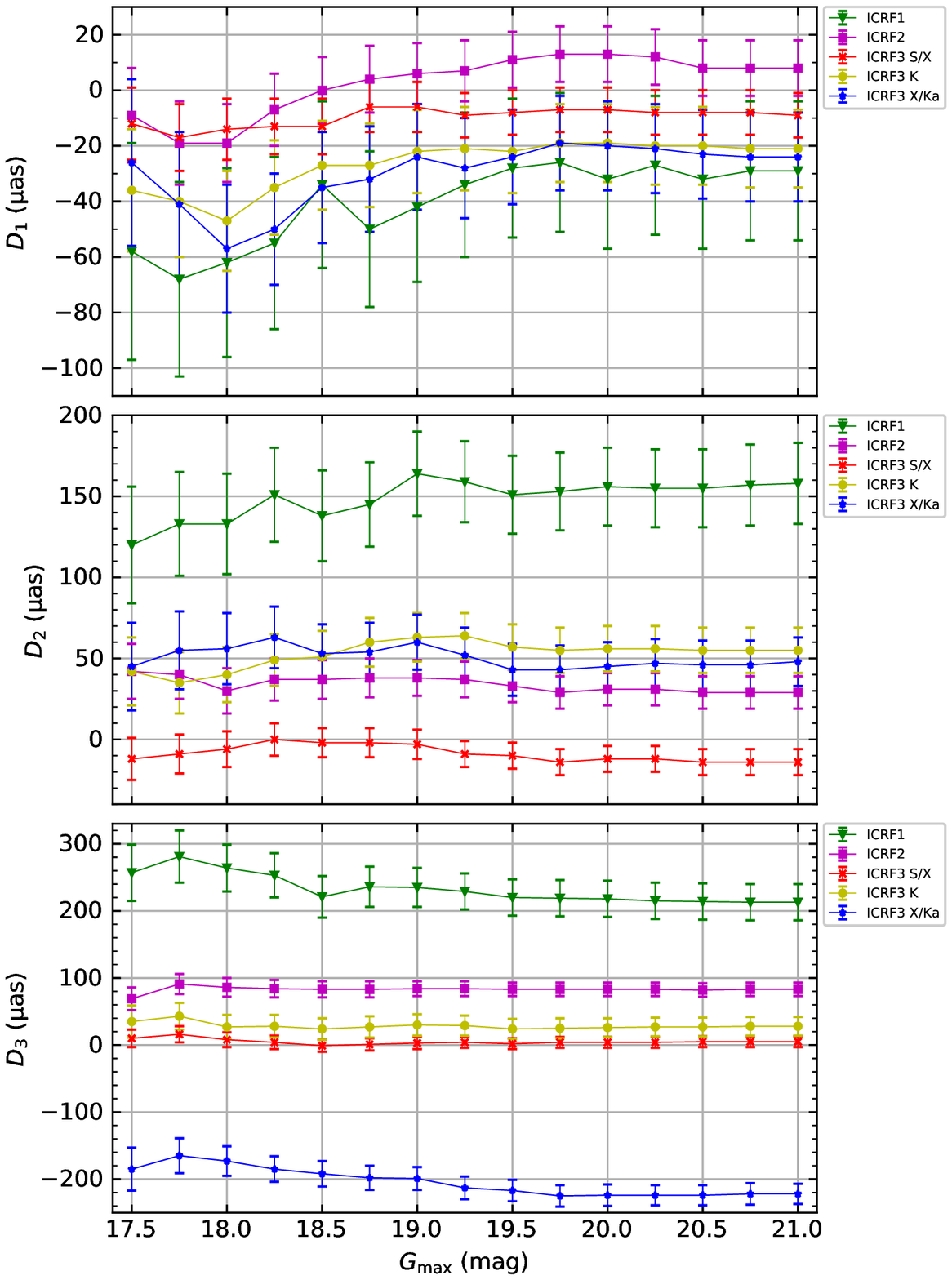}
    \caption[]{\label{fig:glide-on-g}
        Glide parameters of the ICRF positions with respect to the \textit{Gaia} DR2 position as a function of the limiting $G$-magnitude of {\it Gaia} sample.
        The error bar indicates the confidence interval of 1-$\sigma$.
    }
\end{figure}

\begin{figure}[hbtp]
    \centering
    \includegraphics[width=\hsize]{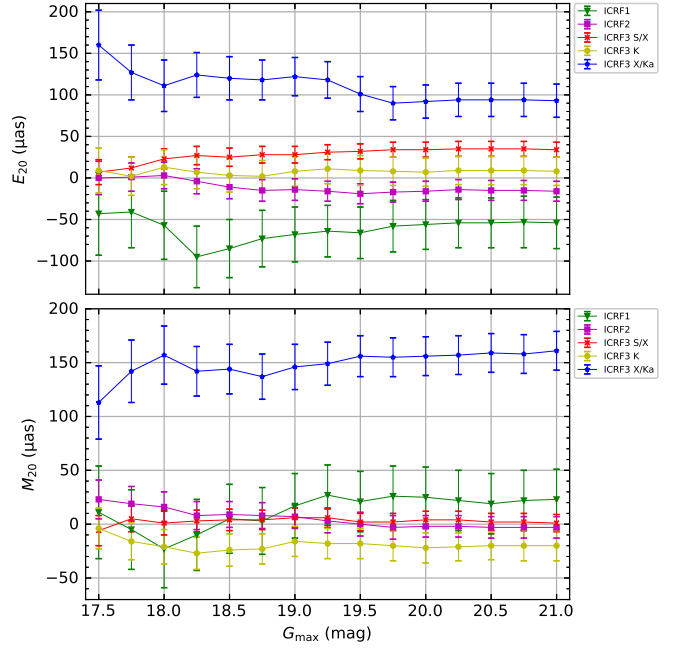}
    \caption[]{\label{fig:quadrupole-on-g}
        Quadrupole parameters $E_{20}$ ($top$) and $M_{20}$ ($bottom$) of the ICRF positions with respect to the \textit{Gaia} DR2 position as a function of the limiting $G$-magnitude of {\it Gaia} sample.
        The error bar indicates the confidence interval of 1-$\sigma$.
    }
\end{figure}

The estimation of rotation, glide, and quadrupole parameters are presented in Fig.~\ref{fig:rotation-on-g}-\ref{fig:quadrupole-on-g}, respectively.
Only the determination of $E_{20}$ and $M_{20}$ are shown; the other quadrupole parameters yield a similar trend to $E_{20}$ and $M_{20}$ and therefore not plotted for brevity.
The ICRF3 S/X- and K-band catalogs yield a stable estimation of VSH parameters across the limiting $G$-magnitude, while for other ICRF catalogs, these parameters stabilize at $G_{\rm max}~>~18.7$.
The obvious variation in VSH parameters for the ICRF3 X/K-band seen in the region of $G_{max}~<~18.7$ is likely due to the small number of used sources in the fit.
In general, all the VSH parameters of ICRF catalogs with referred to {\it Gaia} subsets agree with their formal uncertainties (1-$\sigma$), regardless of different limits imposed on the $G$-magnitude of the {\it Gaia} sample.
This result hints that the precision of {\it Gaia} data gets worse as one moves toward the faint end but not necessarily the accuracy in terms of systematics.
\subsection{Radio-to-optical offset}     \label{sec:separation}
\begin{figure}[hbtp]
    \centering
    \includegraphics[width=\hsize]{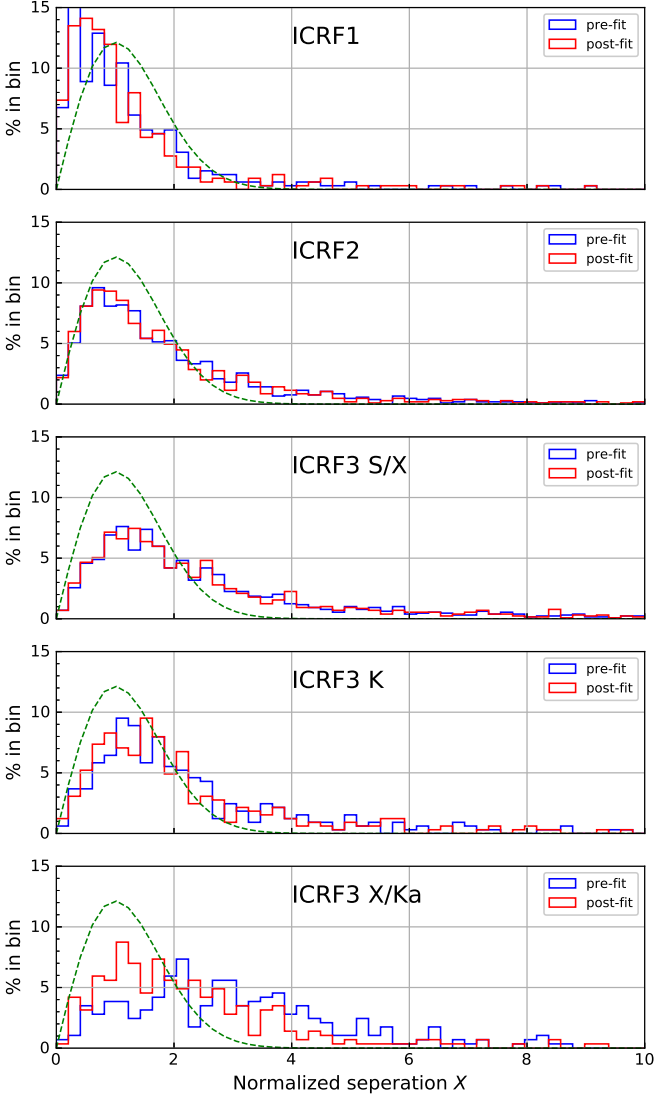}
    \caption[]{\label{fig:nor-sep-dist}
        Distribution of normalized separation between the \textit{Gaia} DR2 and ICRF catalogs before (in blue) and after (in red) adding the adjustments of VSH parameters.
        The green dashed curve indicates a shape of standard Rayleigh distribution.
        
    }
\end{figure}

In this section, we investigated the radio-to-optical offset.
We adjusted the VSH parameters determined in Sect.~\ref{sec:systematics} to the ICRF positions in order to align the radio frame to the {\it Gaia}-CRF2 and recalculated the angular separation and normalized separation.
The distribution of normalized separation after adjustments of VSH parameters together with pre-fit normalized separation is shown in Fig.~\ref{fig:nor-sep-dist}.
The full sample of 2820 sources in the {\it Gaia} DR2 ICRF3-prototype subset was used rather than limited to bright sources ($G<18.7$).
Clearly, the distribution of both pre- and post-fit $X$ deviates from the Rayleigh distribution.
The ICRF1 does not show a large tail but is generally smaller than the prediction from Rayleigh distribution.
It possibly results from the large formal errors in the ICRF1 position.
The ICRF2 fits the Rayleigh distribution well while surprisingly the ICRF3 S/X shows a large outlier rate, the later being similar to the results between \textit{Gaia} DR2 and the prototype solution of ICRF3 \citep{2018A&A...616A..14G}.
Even though the formal error of the ICRF3 has been inflated compared with the prototype solution, it still could not fully explain the radio-to-optic offset.
Comparing the distributions of $X$ for the S/X-, K-, and X/Ka-band, we found that they all deviate from the Rayleigh distribution.
The difference in the shape between pre- and post-fit normalized separation for X/Ka-band is distinct,  
    underlining that the systematics in the X/Ka-band frame will bias the distribution of $X$, which is not seen in other ICRF catalogs.
It suggests that current X/Ka-band position is not reliable enough for statistically studying the radio-to-optical offset unless the systematical deformation in the X/Ka-band frame is corrected.

\begin{figure}[hbtp]
    \centering
    \includegraphics[width=\hsize]{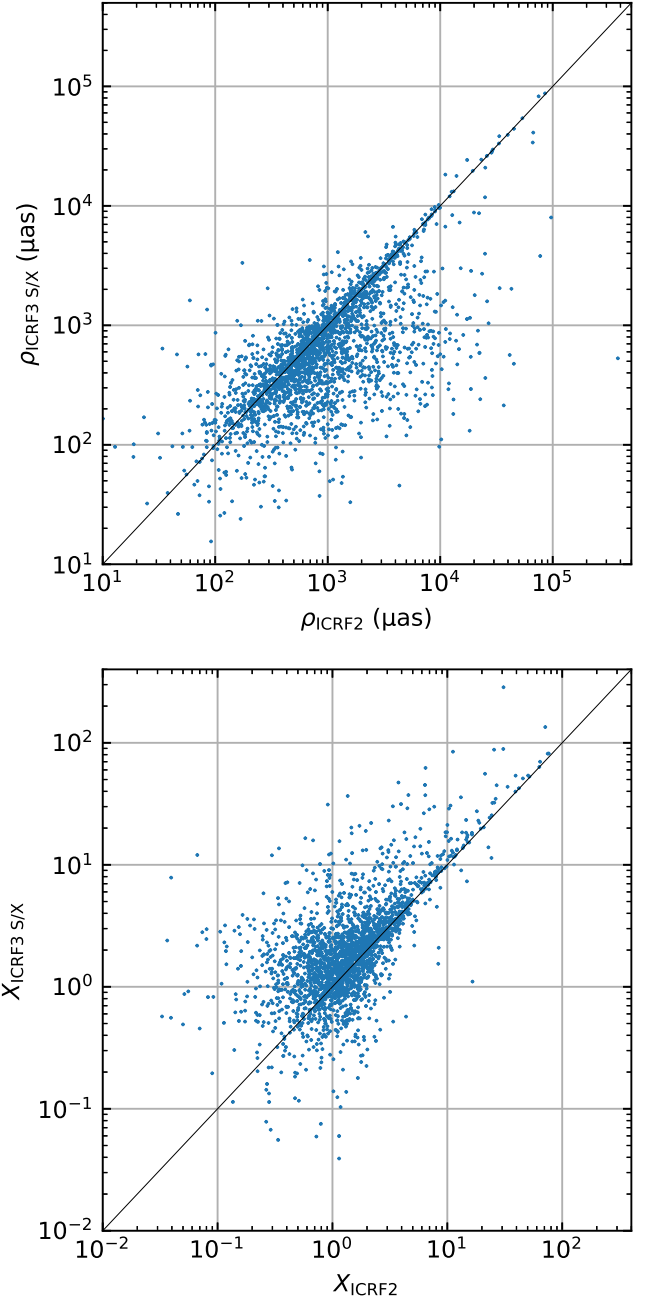}
    \caption[]{\label{fig:sep_icrf2_vs_icrf3}
        Angular separation and normalized separation between the \textit{Gaia} DR2 and ICRF3 S/X catalog for common sources versus those between the \textit{Gaia} DR2 and ICRF2 catalog after adjusting VSH parameters to ICRF positions.
    }
\end{figure}

We further compared the post-fit angular separation and normalized separation of the ICRF2 and ICRF3 S/X with respect to the \textit{Gaia} DR2 (Fig.~\ref{fig:sep_icrf2_vs_icrf3}).
The angular separation between the ICRF2 and \textit{Gaia} DR2 is generally larger than between the ICRF3 S/X and \textit{Gaia} DR2 but the normalized separation is less.
It indicates that some sources with a genuine radio-to-optical offset that was hidden by the large ICRF2 formal error (which leads to a small value of $X$), come out as the precision of the VLBI S/X-band positions improves.
Detailed analysis of these individual radio-to-optical distances, however, is beyond the scope of this work.

\section{Conclusion}     \label{sec:Conclusion}
In this work, we compared the ICRF catalogs of three generations with the \textit{Gaia} DR2 in order to validate the VLBI astrometry.
The new version of the ICRF3 at S/X-band closely agrees with \textit{Gaia} DR2 in terms of the global differences and is nearly zonal (declination-dependent) error-free.
This result supports the high quality of both the ICRF3 and \textit{Gaia}-CRF2.
Compared with its predecessors (the ICRF1 and ICRF2), the ICRF3 S/X-band frame presents a more consistent dependence on the number of observation and a more smooth dependence on the declination, even though one could still find some south-north asymmetry in the positional precision.

The ICRF3 K- and X/Ka-band catalogs show the strength of observations at high frequency.
Even though the number of observation per source is two or three orders smaller than that of S/X-band, the observed positions present a better precision.
However, the possible zonal errors shown in the Sect.~\ref{sec:systematics}, especially dipolar deformation of $\mathrm{\sim-200~\mu as}$ for the X/Ka-band, is most likely due to the weak geometry of the observing network and therefore should be investigated carefully in the future X/Ka-band solution.
Since the observed radio position precision improves rapidly with the number of observations, the K- and X/Ka-band frame could be updated or extended more frequently than the S/X-band frame.

With the improved accuracy of the ICRF3 S/X position, we can see more clearly the radio-to-optical offsets at sub-mas level.
This would benefit the studies of the core-shift effect, but we call the attention for the possible zonal error in the X/Ka-band catalog which should not be omitted.

\begin{acknowledgements}
The authors are grateful to Fran\c{c}ois Mignard for his useful comments and suggestions which improve greatly the manuscript.
The work of N.L. has been supported by China Scholarship Council (CSC) for the joint doctor training program at Department SYRTE (Syst\`{e}mes de R\'{e}f\'{e}rence Temps-Espace) of the Paris Observatory (File No. 201706190125). 
N.L. is also funded by the National Natural Science Foundation of China (NSFC) under grant No. 11833004.
N.L. is grateful to Jiang Nan for her careful check of the language.
This work has made use of data from the European Space Agency (ESA) mission
{\it Gaia} (\url{https://www.cosmos.esa.int/gaia}), processed by the {\it Gaia} Data Processing and Analysis Consortium 
(DPAC,
\url{https://www.cosmos.esa.int/web/gaia/dpac/consortium}). 
Funding for the DPAC
has been provided by national institutions, in particular the institutions
participating in the {\it Gaia} Multilateral Agreement.
This research has made use of NASA's Astrophysics Data System.
This research also made use of Matplotlib, an open-source 2D graphics package for Python\citep{2007CSE.....9...90H} and Astropy,\footnote{\url{http://www.astropy.org}} a community-developed core Python package for Astronomy \citep{2013A&A...558A..33A, 2018AJ....156..123T}.
\end{acknowledgements}

%
   \bibliographystyle{aa} 
   \bibliography{crf-comparison} 
%



\end{document}